\documentstyle [12pt] {article}
\begin{document}
\setcounter{page}{1}
\def\theequation{\arabic{section}.\arabic{equation}}
\def\theequation{\thesection.\arabic{equation}}
\setcounter{section}{0}

\title{ Nambu--Jona--Lasinio approach to realization of confining medium}

\author{M. Faber\thanks{E--mail: faber@kph.tuwien.ac.at, Tel.:
+43--1--58801--5598, Fax: +43--1--5864203} ,
 A. N. Ivanov\thanks{E--mail: ivanov@kph.tuwien.ac.at,
Tel.: +43--1--58801--5598, Fax: +43--1--5864203}~
{\footnotesize$^{^{\ddag}}$},
N.I. Troitskaya\thanks{Permanent Address:
State Technical University, Department of Theoretical
Physics, 195251 St. Petersburg, Russian Federation}}

\date{}

\maketitle

\begin{center}
{\it Institut f\"ur Kernphysik, Technische Universit\"at Wien, \\
Wiedner Hauptstr. 8-10, A-1040 Vienna, Austria}
\end{center}

\vskip1.0truecm
\begin{center}
\begin{abstract}
The mechanism of a confining medium is investigated within the
Nambu--Jona--Lasinio (NJL) approach. It is shown that a confining
medium can be realized in the bosonized phase of the NJL model due to
vacuum fluctuations of both fermion and Higgs (scalar fermion--antifermion
collective excitation) fields. In such an approach there is no need to
introduce Dirac strings.
\end{abstract}
\end{center}

\begin{center}
PACS: 11.15.Tk, 12.39.Ki, 12.60.Rc, 14.80.Cp.\\
\noindent Keywords: phenomenological quark model, technifermions,
Higgs field, confining medium, confinement, composed states, linear
potential
\end{center}
\vskip1.0truecm

\newpage

\section{Introduction}
\setcounter{equation}{0}

\hspace{0.2in} Mechanism of confining medium suggested in Abelian Higgs
model by Narnhofer and Thirring [1] and in a dual QCD by Baker, Ball and
Zachariasen [2] can be rather useful for the realization of quark confinement
 by a way avoiding the problem of the inclusion of Dirac strings. The
dielectric constant of a confining medium $\varepsilon (k^2)$ vanishes
at large distances like $\varepsilon (k^2) \sim k^2$ which leads to a
linearly rising interquark potential realizing confinement of quarks [3].

The approaches to confining medium [1,2] are based on the similarity between
dual superconductivity and the Higgs mechanism inducing non--perturbative
vacuum with properties of superconductor. The Nambu--Jona--Lasinio (NJL)
 model [4--10], being a relativistic extension of the BCS
(Bardeen--Cooper--Schrieffer) theory of superconductivity [11], gives the
 alternative mechanism of the realization of a superconducting
non--perturbative vacuum. Recently in Ref.[10] we have investigated the
mechanism of quark confinement in the Abelian monopole NJL model with dual
Dirac strings. In this model quarks and antiquarks are classical particles
joined to the ends of dual Dirac strings, while monopoles are quantum massless
 fermion fields which become massive due to monopole--antimonopole
condensation induced by four--monopole interaction. The
monopole--antimonopole condensation
 accompanies itself the creation of the monopole--antimonopole scalar and
dual--vector collective excitations. The ground state of the dual--vector
field induced by the interaction with a dual Dirac string has the shape of
the Abrikosov flux line. The latter leads to linearly rising potential and
confinement of quarks and antiquarks joined to the ends of a dual Dirac string.

This paper is to apply the NJL model analoguous the Abelian monopole NJL model
 [10] to the realization of confining medium and consider the way avoiding
the inclusion of dual Dirac strings. Since we do not include dual Dirac
strings, the quarks and antiquarks, confinement of which we are investigating,
 are described now by an external field $\psi(x)$ related to an external
electric current $j_{\mu}(x) = \bar{\psi}(x) \gamma_{\mu}\psi(x)$. Then,
due to the absence of dual Dirac strings instead of the monopole fermion
fields we use fermion fields like technifermions introduced in the
technicolour approach [12] to the standard electroweak model for the
description of the appearance of
 the $W$ and $Z$--boson masses without Higgs mechanism. The NJL model in our
consideration is an Abelian one, and we need to introduce only one sort of
technifermions. For the definitness we would call them electroquarks and
define by the field $\chi(x)$. The electroquarks are massless and acquire
the mass due to strong local four--electroquark interactions.

The Lagrangian of the starting system should read [10]
\begin{eqnarray}\label{label1.1}
{\cal L}(x)&=&\bar{\chi}(x)\,i\,\gamma^{\mu}\,\partial_{\mu}\chi(x) +
G\, [\bar{\chi}(x)\,\chi(x)]^2 \nonumber\\
&&-G_1\, [\bar{\chi}(x)\,\gamma_{\mu}\,\chi(x) + j_{\mu}(x)]\,
[\bar{\chi}(x)\,\gamma^{\mu}\,\chi(x) + j^{\mu}(x)]\,,
\end{eqnarray}
where $G$ and $G_1$ are positive phenomenological coupling constants that
 we fix below.

The Lagrangian Eq.(\ref{label1.1}) is invariant under $U(1)$ group. Due to
 strong attraction in the $\bar{\chi}\chi$--channels produced by the local
 four--electroquark interaction Eq.(\ref{label1.1}) the $\chi$--fields
become unstable under $\bar{\chi}\chi$ condensation, i.e.
 $<\bar{\chi}\chi>\not=0$. Indeed, in the condensed phase the energy
of the ground state of the electroquark fields is negative, i.e.
\begin{eqnarray}\label{label1.2}
{\cal W} = - <{\cal L}_{\chi}(x)> = - \Bigg(\frac{3}{4}\,G + G_1\Bigg)\,
[<\bar{\chi}\chi>]^2 < 0,
\end{eqnarray}
whereas in the non--condensed phase, when $<\bar{\chi}\chi>=0$, we have
${\cal W} = 0$. This means that the condensed phase is much more advantageous
to the electroquark system, described by the Lagrangian (\ref{label1.1}).
 The electroquark fields become condensed without breaking of $U(1)$
symmetry as well.

In the condensed phase the electroquark fields acquire a mass $M$
satisfying the gap--equation [4--10]
\begin{eqnarray}\label{label1.3}
&&M=-2 G <\bar{\chi}(0) \chi(0)> = \frac{G M}{2\pi^2}J_1(M) =\nonumber\\
&&=\frac{G M}{2\pi^2}\int \frac{d^4k}{\pi^2 i}\frac{1}{M^2 - k^2} =
\frac{G M}{2 \pi^2} \Bigg[\Lambda^2 -  M^2
{\ell n}\Bigg(1 + \frac{\Lambda^2}{M^2}\Bigg)\Bigg],
\end{eqnarray}
where $\Lambda$ is the ultra--violet cut--off.

The condensation of the electroquark fields accompanies the creation of
$\bar{\chi}\chi$ collective excitations with the quantum numbers of a scalar
 Higgs meson field $\rho$ and a vector field $A_{\mu}$.

The main aim of the approach is to show that in the tree approximation for
the $A_{\mu}$--field and after the integration over the electroquark and
the scalar fields the effective Lagrangian of the $A_{\mu}$--field acquires
the form [1,2]
\begin{eqnarray}\label{label1.4}
{\cal L}_{\rm eff}[A(x)]\,=\,-\,\frac{1}{4}\,F_{\mu\nu}(x)\,\varepsilon\,
(\Box)\,F^{\mu\nu}(x) - j_{\mu}(x) \,A^{\mu}(x)\,,
\end{eqnarray}
where $F_{\mu\nu}(x) = \partial_{\mu} A_{\nu}(x) - \partial_{\nu} A_{\mu}(x)$
and $\varepsilon(\Box)$ is the operator of the dielectric constant which in
the momentum representation $\varepsilon(k^2)$ vanishes at large distances,
i.e. $\varepsilon (k^2) \sim k^2$ at $k^2 \to 0$. Due to the tree
$A_{\mu}$--field approximation integrating over the electroquark and the
scalar fields we can keep only the terms proportional to $F_{\mu\nu}(x)\,
F^{\mu\nu}(x)$ and $A_{\mu}(x)\,A^{\mu}(x)$.

The paper is organized as follows. In Sect.~2 we derive the effective
Lagrangian for collective excitations of electroquarks in one--loop
approximation keeping only the main divergent contibutions and leading
order in long--wavelength
expansion. That is in accordance with the standard approximation accepted in
the NJL model approaches. We show that such a way does lead to the realization
of a non--trivial medium. In Sect.~3 we take into account convergent
contributions of the one--electroquark loop diagrams that means the step
beyond the standard approximation. This has led to the effective Lagrangian
for the vector collective excitations coupled to scalar collective
excitations.
Integrating out the scalar collective excitations, since they are much heavier
than the vector ones, we arrive at the effective Lagrangian for the massless
vector collective excitation field in a non--trivial medium with the
dielectric constant vanishes at large distances. In Sect.~4 we show that the
external $\psi$--quarks couple to each other and the medium via the exchange
of the massless vector collective excitations become confined due to a
linearly rising potential induced by the medium. In Sect.~5 we discuss the
obtained
results.

\section{Effective Lagrangian. Standard approximation}
\setcounter{equation}{0}

Following [5--10] we define the effective Lagrangian of the $\rho$ and
$A_{\mu}$ fields as follows
\begin{eqnarray}\label{label2.1}
{\cal L}_{\rm eff}(x)\,=\,{\tilde{\cal L}}_{\rm eff}(x) - \frac{\kappa^2}{4G}
 \rho^2(x) + \frac{g^2}{4 G_1} A_{\mu}(x) \,A^{\mu}(x)  - j_{\mu}(x) \,
A^{\mu}(x)\,,
\end{eqnarray}
where
\begin{eqnarray}\label{label2.2}
{\tilde{\cal L}}_{\rm eff}(x)\,=\,-\,i\,
\Bigg<\,x\Bigg|{\ell n}\frac{{\rm Det}(i\,\hat{\partial}\,-\,M\,+\,\Phi)}
{{\rm Det}(i\,\hat{\partial}\,-\,M)}\Bigg|x\,\Bigg>\,.
\end{eqnarray}
Here we have denoted $\Phi = - g \gamma^{\mu} A_{\mu} - \kappa \sigma$, and
$\sigma = \rho - M/\kappa$, where $g$ is the electric charge of the
electroquark field that we fix below. In the tree approximation the
$\sigma$--field has a vanishing vacuum expectation value (v.e.v.), i.e.
 $<\sigma>\,=\,0$, whereas the v.e.v. of the $\rho$--field does not vanish,
 i.e. $<\rho>\,=\,M/\kappa\,\not=\,0$.

The effective Lagrangian ${\tilde{\cal L}}_{\rm eff}(x)$ can be represented
 by an infinite series
\begin{eqnarray}\label{label2.3}
{\tilde{\cal L}}_{\rm eff}(x) = \sum_{n = 1}^{\infty}\frac{i}{n}\,
{\rm tr}_{\,\rm L}\Bigg< x \Bigg|\Bigg(\frac{1}{M\,-\,i\,\hat{\partial}}\,
\Phi\Bigg)^n \Bigg|x\,\Bigg>\,=\,\sum_{n = 1}^{\infty}
{\tilde{\cal L}}^{(n)}_{\rm eff}(x) \,.
\end{eqnarray}
The subscript ${\rm L}$ means the computation of the trace over Dirac
matrices. The effective Lagrangian ${\tilde{\cal L}}^{\,(n)}_{\rm eff}(x)$
is given
by [5--10,13]
\begin{eqnarray}\label{label2.4}
\tilde{{\cal L}}^{(n)}_{\rm eff}\,(x)&=&\int\,
\prod_{\ell =\,1}^{n\,-\,1}
\frac{d^4\,x_{\,\ell}\,d^4\,k_{\,\ell}}{(2\,\pi)^{\,4}}\,
e^{\,-\,i\,k_{\,1}\cdot x_1\,-\ldots -i\,k_{\,n}\cdot x\,}\,
\left(-\,\frac{1}{n}\,\frac{1}{16\,\pi^2}\right)\,
\int\frac{d^4\,k}{\pi^2\,i}\nonumber\\
&&\times{\rm tr}_{\,\rm L}\,\Bigg\{\,\frac{1}{M\,-\,\hat k}\,
\Phi\,(x_1)\,\frac{1}{M\,-\,\hat k\,-\,{\hat k}_{\,1}}\,\Phi\,(x_2)\,\ldots
\nonumber\\
&& \times \ldots \Phi\,(x_{\,n\,-\,1})\,
\frac{1}{M\,-\,\hat k\,-\,{\hat k}_{\,1}\,-\ldots -\,{\hat
k}_{\,n\,-\,1}}\,\Phi\,(x)\,\Bigg\}
\end{eqnarray}
at $k_1 + k_2 + \ldots + k_n = 0$. The r.h.s. of (\ref{label2.4}) describes
the one--electroquark loop diagram with $n$--vertices. The
one--electroquark loop diagrams with two vertices $(n = 2)$ determine the
kinetic term of the $\sigma$--field and give the contribution to the
kinetic term of the $A_{\mu}$--field, while the diagrams with ($n\ge 3$)
describe the vertices of interactions of the $\sigma$ and $A_{\mu}$ fields.
According to the NJL prescription the effective Lagrangian $\tilde{{\cal
L}}_{\rm eff}(x)$ should be  defined by the set of divergent
one--electroquark loop diagrams with
 $n = 1,2,3$ and $4$ vertices [5--10]. The compution of these diagrams we
perform at leading order in the long--wavelength expansion approximation
accepted in the NJL model [5--10] when gradients of $A_{\mu}$ and $\sigma$
fields are slowly varying fields. This approximation is fairly good
established for sufficiently heavy electorquark fields $\chi(x)$ that we
assume following the technicolour extension of the standard electroweak
model [12]. As a result we get
\begin{eqnarray}\label{label2.5}
&&{\cal L}_{\rm eff}(x) = \nonumber\\
&&=- \frac{g^2}{48\pi^2}\,J_2(M)\, F_{\mu\nu}(x) F^{\mu\nu}(x) +
\Bigg\{\frac{g^2}{4 G_1}-\frac{g^2}{16 \pi^2} [J_1(M) + M^2
J_2(M)]\Bigg\}\,A_{\mu}(x)\,A^{\mu}(x) \nonumber\\
&& +
\frac{1}{2}\frac{\kappa^2}{8\pi^2}J_2(M)\partial_{\mu}\sigma(x)\,\partial^{\mu}\
sigma(x) - M\,\Bigg[\frac{\kappa}{2 G} - \frac{\kappa}{4 \pi^2}
J_1(M)\Bigg]\,\sigma(x)\nonumber\\
&&+ \frac{1}{2}\Bigg[- \frac{\kappa^2}{2 G} + \frac{\kappa^2}{4 \pi^2}J_1(M)
 - 4 M \frac{\kappa^2}{8 \pi^2} J_2(M)\Bigg]\,\sigma^2(x) - 2
M\,\kappa\,\frac{\kappa^2}{8 \pi^2} J_2(M)\,\sigma^3(x)\nonumber\\
&&- \frac{1}{2} \kappa^2\,\frac{\kappa^2}{8 \pi^2}\,J_2(M)\,\sigma^4(x) -
j_{\mu}(x) \,A^{\mu}(x).
\end{eqnarray}
In order to get correct factors of the kinetic terms of the $\sigma$ and
$A_{\mu}$ fields we have to set [5--10]
\begin{equation}\label{label2.6}
\frac{g^2}{12 \pi^2} J_2(M) = 1 \quad,\quad \frac{\kappa^2}{8 \pi^2}
 J_2(M) = 1 ,
\end{equation}
that arranges the relation $\kappa^2 = 2g^2/3$ [5--10], and $J_2(M)$ is a
logarithmically divergent integral defined by
\begin{eqnarray}\label{label2.7}
J_2(M) = \int  \frac{d^4k}{\pi^2 i} \frac{1}{(M^2 - k^2)^2} =
{\ell n}\Bigg(1 + \frac{\Lambda^2}{M^2}\Bigg) -
\frac{\Lambda^2}{M^2 + \Lambda^2}\,.
\end{eqnarray}
The relations (\ref{label2.6}) can be represented in a more comprehensible
 way in terms of constraints for the renormalization constants of the
wave--functions of the $A_{\mu}$ and $\sigma$ fields. For this aim we
rewrite the Lagrangian (\ref{label2.5}) as follows
\begin{eqnarray}\label{label2.8}
&&{\cal L}_{\rm eff}(x) =\\
&&= - \frac{1}{4} (1 - Z^{\chi}_{\rm A}) F_{\mu\nu}(x)
F^{\mu\nu}(x)+\frac{1}{2}(1 - Z^{\chi}_{\sigma})\partial_{\mu}\sigma(x)
\partial^{\mu}\sigma(x)+
\ldots\,,\nonumber
\end{eqnarray}
where
\begin{equation}\label{label2.9}
Z^{\chi}_{\rm A} = 1 - \frac{g^2}{12\,\pi^2}\,J_2(M)\quad,\quad
Z^{\chi}_{\sigma} = 1 - \frac{\kappa^2}{8\,\pi^2}\,J_2(M)
\end{equation}
are the renormalization constants of the wave--functions of the $A_{\mu}$
and $\sigma$ fields procreated by vacuum fluctuations of the electroquark
fields.
In Eq.(\ref{label2.8}) we have kept only kinetic terms of the $A_{\mu}$ and
$\sigma$ fields, other terms are irrelevant at the moment. Since $A_{\mu}$
and $\sigma$ are bound $\bar{\chi}\chi$--states, the renormalization
constants $Z^{\chi}_{\rm A}$ and $Z^{\chi}_{\sigma}$ should vanish due to
the so--called compositeness condition $Z^{\chi}_{\rm A} =
Z^{\chi}_{\sigma} = 0$ [14], i.e.
\begin{equation}\label{label2.10}
Z^{\chi}_{\rm A}=1- \frac{g^2}{12\,\pi^2}\,J_2(M)=0\quad,\quad
Z^{\chi}_{\sigma}=1-\frac{\kappa^2}{8\,\pi^2}\,J_2(M)=0.
\end{equation}
The compositeness condition can be applied to bound states in
non--perturbative quantum field theory [14]. As a result we arrive at
Eq.(\ref{label2.6}). Picking up Eq.(\ref{label1.3}) and Eq.(\ref{label2.6})
we bring up the effective Lagrangian (\ref{label2.5}) to the form
\begin{eqnarray}\label{label2.11}
{\cal L}_{\rm eff}(x)&=&-\frac{1}{4} F_{\mu\nu}(x) F^{\mu\nu}(x)   +
\,\frac{1}{2} M^2_A A_{\mu}(x) A^{\mu}(x) - j^{\mu}(x) A_{\mu}(x)
\nonumber\\
&+&\frac{1}{2} \partial_{\mu} \sigma(x) \partial^{\mu} \sigma(x) -
\frac{1}{2} M^2_{\sigma} \sigma^2(x)
\Bigg[1 + \kappa \frac{\sigma(x)}{M_{\sigma}}\Bigg]^2 -
 j_{\mu}(x) \,A^{\mu}(x),
\end{eqnarray}
where $M_{\sigma} = 2 M$ is the mass of the $\sigma$--field and
\begin{eqnarray}\label{label2.12}
M^2_A = \frac{g^2}{2 G_1}-\frac{g^2}{8 \pi^2} [J_1(M) + M^2 J_2(M)]
\end{eqnarray}
is the squared mass of the $A_{\mu}$--field. Without loss of generality one
can assume that $M_A \ll M_{\sigma} = 2\,M$. This should allow one to
integrate over the heavy scalar collective excitations and derive an
effective Lagrangian only for the vector ones.

The Lagrangian Eq.(\ref{label2.11}) describes the effective Lagrangian of
the collective excitations ($\bar{\chi}\chi$--bound states) derived within
the standard procedure of the NJL approach [4--10], i.e. at leading order
in long--wavelength expansion of the one--electroquark loop diagrams.

The effective Lagrangian describing the propagation of the $A_{\mu}$--field
in a dielectric medium should take the form
Eq.(\ref{label1.4}). If it is a confining medium, the Fourier transform of
a dielectric constant $\varepsilon(k^2)$ should vanish at $k^2\to 0$, i.e.
$\varepsilon(k^2)\to 0$ at $k^2\to 0$ [1,2]. It is seen that the Lagrangian
Eq.(\ref{label2.10}) describes a dielectric medium with $\varepsilon = 1$.
This should imply that {\it the realization of the confining medium in the
NJL approach goes beyond the standard procedure of the derivation of the
effective Lagrangian describing collective excitations}. The simplest
extension of the
NJL prescription is to take into account {\it the contributions of
convergent electroquark  loop diagrams and the Higgs field loops}. As has
been shown in Ref.[9] contributions of convergent electroquark loop
diagrams play an
important role for processes of low--energy interactions of low--lying hadrons.

\section{Effective Lagrangian for confining medium}
\setcounter{equation}{0}

Thus, in order to obtain non--trivial contributions to the dielectric
constant we should leave a standard approximation for the derivation of the
effective Lagrangians in the NJL models.

Since we are interested in the kinetc terms of the $A_{\mu}$--field,
the simplest step leading beyond the standard approximation is to take into
account convergent contributions of the one--electroquark loop diagrams to
the kinetic term of the vector field $A_{\mu}$. The exact calculation of
the electroquark loop diagram with two--vector vertices alters the
effective Lagrangian (\ref{label2.11}) as follows
\begin{eqnarray}\label{label3.1}
{\cal L}_{\rm eff}(x)&=&-\frac{1}{4} F_{\mu\nu}(x) \varepsilon(\Box)\,
F^{\mu\nu}(x)  + \frac{1}{2} M^2_A A_{\mu}(x) A^{\mu}(x) -
j^{\mu}(x) A_{\mu}(x) \nonumber\\
&&+\frac{1}{2} \partial_{\mu} \sigma(x)   \partial^{\mu} \sigma(x) -
 \frac{1}{2} M^2_{\sigma} \sigma^2(x) \Bigg[1 + \kappa
\frac{\sigma(x)}{M_{\sigma}}\Bigg]^2\,,
\end{eqnarray}
where $\varepsilon(\Box)$ is given by
\begin{eqnarray}\label{label3.2}
\varepsilon(\Box) = 1 - \frac{g^2}{2 \pi^2} \int^1_0
d\eta\,\eta(1-\eta){\ell n}\Bigg[1 + \frac{\Box}{M^2}\,
\eta(1-\eta)\Bigg]\,,
\end{eqnarray}
the Fourier transform of which yields
\begin{eqnarray}\label{label3.3}
\varepsilon(k^2) = 1 - \frac{g^2}{2 \pi^2}\int^1_0
d\eta\,\eta(1-\eta){\ell n}\Bigg[1 - \frac{k^2}{M^2}\,\eta(1-\eta)\Bigg]\,.
\end{eqnarray}
We have dropped the convergent contributions of the electroquark loop
diagrams to the other terms of the effective Lagrangian
Eq.(\ref{label3.1}), since they are less important for the problem under
consideration.

The other non--trivial contributions to the kinetic term of the
$A_{\mu}$--field come from the convergent one--electroquark loop diagrams
inducing the interactions between $\sigma$ and $A_{\mu}$ fields. They read
\begin{eqnarray}\label{label3.4}
\delta\,{\cal L}_{\rm eff}(x) &=& M\frac{\kappa
g^2}{8\pi^2}\sigma(x)\Bigg[1+\kappa\frac{\sigma(x)}{2M}\Bigg]
A_{\mu}(x)A^{\mu}(x) + \nonumber\\
&&+ \frac{g^2}{24\pi^2}{\ell n}\Bigg[1+\kappa\frac{\sigma(x)}{M}\Bigg]
F_{\mu\nu}(x) F^{\mu\nu}(x)\,.
\end{eqnarray}
This is the most general interaction of two vector mesons with scalar
fields derived in leading order of the expansion in powers of gradients of
the $\sigma$--field, $\partial_{\mu}\sigma(x)$, which are slowly varying
fields.

By appending the interaction Eq.(\ref{label3.4}) to the  Lagrangian
(\ref{label3.1}) we arrive at the following effective Lagrangian
\begin{eqnarray}\label{label3.5}
{\cal L}_{\rm eff}(x) ={\cal L}_{\rm eff}[A(x),\sigma(x)] + \frac{1}{2}
\partial_{\mu} \sigma(x) \partial^{\mu} \sigma(x) - \frac{1}{2}
M^2_{\sigma} \sigma^2(x) \Bigg[1 + \kappa
\frac{\sigma(x)}{M_{\sigma}}\Bigg]^2,
\end{eqnarray}
where we have denoted
\begin{eqnarray}\label{label3.6}
{\cal L}_{\rm eff}[A(x),\sigma(x)] = - \frac{1}{4} F_{\mu\nu}(x)
\varepsilon(\Box,\sigma) F^{\mu\nu}(x) +
 \frac{1}{2} M^2_A (\sigma) A_{\mu}(x) A^{\mu}(x) - j^{\mu}(x) A_{\mu}(x).
\end{eqnarray}
The dielectric constant $\varepsilon(\Box, \sigma)$ reads
\begin{eqnarray}\label{label3.7}
\varepsilon(\Box, \sigma)&=& 1 - \frac{g^2}{6\pi^2}{\ell
n}\Bigg[1+\kappa\frac{\sigma(x)}{M}\Bigg]- \nonumber\\
&-&\frac{g^2}{2 \pi^2}\int^1_0d\eta\,\eta(1-\eta){\ell n}\Bigg[1 +
\frac{\Box}{M^2}\,\eta(1-\eta)\Bigg]\,,
\end{eqnarray}
and $M^2_A(\sigma)$ is defined
\begin{eqnarray}\label{label3.8}
M^2_A(\sigma) = \frac{g^2}{2 G_1}-\frac{g^2}{8\pi^2}[J_1(M)+M^2J_2(M)] +
\frac{\kappa g^2}{4 \pi^2}M \sigma(x)
\Bigg[1+\kappa\frac{\sigma(x)}{2M}\Bigg].
\end{eqnarray}
It is seen that a dielectric constant has become a functional of the
$\sigma$--field. Thereby, one can expect a substantial influence of the
vacuum fluctuations of the $\sigma$--field on the dielectric constant.
Since the $\sigma$--field is much heavier than the $A_{\mu}$--field, i.e.
$M_{\sigma} = 2\,M\gg M_A$, at low energies only vacuum fluctuations of the
$\sigma$--field are important. Therefore, in order to pick up the
contribution of vacuum fluctuations of the $\sigma$--field we have to
integrate it out. The effective Lagrangian ${\cal L}_{\rm eff}[A_{\mu}(x)]$
can be defined [13]
\begin{eqnarray}\label{label3.9}
e^{i\int  \displaystyle d^4x{\cal L}_{\rm eff}[A_{\mu}(x)]}=
\int {\cal D}\sigma \,e^{i\int  \displaystyle  d^4x\,
\{{\cal L}_{\rm eff}[A_{\mu}(x),\sigma(x)] +
{\cal L}_{\rm eff}[\sigma(x)]\}},
\end{eqnarray}
where ${\cal L}_{\rm eff}[\sigma(x)]$ reads
\begin{eqnarray}\label{label3.10}
{\cal L}_{\rm eff}[\sigma(x)] = \frac{1}{2} \partial_{\mu} \sigma(x)
\partial^{\mu} \sigma(x) - \frac{1}{2} M^2_{\sigma} \sigma^2(x)
\Bigg[1 + \kappa \frac{\sigma(x)}{M_{\sigma}}\Bigg]^2 -
V_{\rm eff}[\sigma(x)],
\end{eqnarray}
where $V_{\rm eff}[\sigma(x)]$ describes the contribution of the convergent
electroquark loops yielding self--interactions of the $\sigma$--field at
leading order of the expansion in powers of the gradients of the
$\sigma$--field. One can expect that $V_{\rm eff}[\sigma(x)]\sim
\sigma^6(x) + \ldots$. The integrals over the $\sigma$--field can be
normalized by the condition
\begin{eqnarray}\label{label3.11}
\int {\cal D}\sigma \,e^{i\int  \displaystyle  d^4x\,
{\cal L}_{\rm eff}[\sigma(x)]}= 1.
\end{eqnarray}
Of course, the exact integration over the $\sigma$--field cannot be done and
we should develop an approximate scheme.

For this aim it is convenient to rewrite Eq.(\ref{label3.9}) as follows
\begin{eqnarray}\label{label3.12}
&&e^{i\int  \displaystyle  d^4x{\cal L}_{\rm eff}[A_{\mu}(x)]}=
\nonumber\\
&&= e^{i\int  \displaystyle  d^4x \Big[-\frac{1}{4}\,F_{\mu\nu}(x)
\varepsilon(\Box) F^{\mu\nu}(x) + \frac{1}{2} M^2_A A_{\mu}(x) A^{\mu}(x) -
j_{\mu}(x) A^{\mu}(x)\Big]}\nonumber\\
&&\int {\cal D}\sigma \,e^{i\int  \displaystyle   d^4x\,\Big\{ -
\frac{1}{4} F_{\mu\nu}(x)[\varepsilon(\Box,\sigma) - \varepsilon(\Box)]
F^{\mu\nu}(x) +  \frac{1}{2} [M^2_A (\sigma) - M^2_A]A_{\mu}(x)
A^{\mu}(x)\Big\}}
\nonumber\\
&&e^{i\int  \displaystyle  d^4x\,{\cal L}_{\rm eff}[\sigma(x)]},
\end{eqnarray}
Recall that the effective Lagrangian Eq.(\ref{label3.6}) has been
calculated keeping the quadratic terms in the $A_{\mu}$--field. This means
that in the integrand we can expand the exponential keeping only quadratic
terms in the $A_{\mu}$--field expansion
\begin{eqnarray}\label{label3.13}
&&e^{i\int  \displaystyle  d^4x\, {\cal L}_{\rm eff}[A_{\mu}(x)]}=
 \nonumber\\
&&= e^{i\int  \displaystyle  d^4x \Big[-\frac{1}{4}\,F_{\mu\nu}(x)
\varepsilon(\Box) F^{\mu\nu}(x) + \frac{1}{2} M^2_A A_{\mu}(x) A^{\mu}(x) -
j_{\mu}(x) A^{\mu}(x)\Big]}\nonumber\\
&&\int {\cal D}\sigma \,\Big\{1 - \frac{1}{4} i\int d^4x
F_{\mu\nu}(x)[\varepsilon(\Box,\sigma) - \varepsilon(\Box)] F^{\mu\nu}(x)
\nonumber\\
&&\hspace{1in} + \frac{1}{2}i\int d^4x [M^2_A (\sigma) - M^2_A] A_{\mu}(x)
A^{\mu}(x)\Big\}e^{i\int  \displaystyle  d^4x\,{\cal L}_{\rm
eff}[\sigma(x)]} \nonumber\\
&&= e^{i\int  \displaystyle  d^4x \Big[-\frac{1}{4}\,F_{\mu\nu}(x)
\varepsilon(\Box) F^{\mu\nu}(x) + \frac{1}{2} M^2_A A_{\mu}(x) A^{\mu}(x) -
j_{\mu}(x) A^{\mu}(x)\Big]}\nonumber\\
&&\Big\{1  - \frac{1}{4}i\int d^4x\, F_{\mu\nu}(x)<\varepsilon(\Box,\sigma)
- \varepsilon(\Box)> F^{\mu\nu}(x) \nonumber\\
&&\hspace{1in}+ \frac{1}{2}i\int d^4x\,<M^2_A (\sigma) - M^2_A> A_{\mu}(x)
A^{\mu}(x)\Big\}\nonumber\\
&&\simeq e^{i\int  \displaystyle  d^4x \Big[-\frac{1}{4}\,F_{\mu\nu}(x)
<\varepsilon(\Box,\sigma)> F^{\mu\nu}(x) +} \nonumber\\
&&\hspace{1in}+ \frac{1}{2}< M^2_A(\sigma)> A_{\mu}(x) A^{\mu}(x) -
 j_{\mu}(x) A^{\mu}(x)\Big].
\end{eqnarray}
Thus, up to the quadratic terms in the $A_{\mu}$--field expansion the
effective Lagrangian ${\cal L}_{\rm eff}[A_{\mu}(x)]$ is given by
\begin{eqnarray}\label{label3.14}
{\cal L}_{\rm eff}[A(x)]&=&- \frac{1}{4} F_{\mu\nu}(x)
<\varepsilon(\Box,\sigma)> F^{\mu\nu}(x) \nonumber\\
&&+\frac{1}{2} <M^2_A (\sigma)> A_{\mu}(x) A^{\mu}(x) - j^{\mu}(x) A_{\mu}(x).
\end{eqnarray}
The derivation of the Lagrangian Eq.(\ref{label3.14}) has been performed in
the $A_{\mu}$--field tree approximation. In this case the operators
$F_{\mu\nu}(x)F^{\mu\nu}(x)$ and $A_{\mu}(x)A^{\mu}(x)$ are not affected by
the contributions of higher powers in the $A_{\mu}$--field expansion.
Therefore, in the $A_{\mu}$--field tree approximation the justification of
the validity of the derivation of the Lagrangian Eq.(\ref{label3.14}) does
not need the smallness of higher power terms in the $A_{\mu}$--field
expansion.

The expectation values $<\varepsilon(\Box,\sigma)>$ and $<M^2_A (\sigma)>$ read
\begin{eqnarray}\label{label3.15}
&&<\varepsilon(\Box,\sigma)>= \int {\cal D}\sigma
\,\varepsilon(\Box,\sigma)\,\exp{i\int d^4z\,{\cal L}_{\rm
eff}[\sigma(z)]}=\nonumber\\
&&= 1 -\Bigg<\frac{g^2}{6\pi^2}{\ell
n}\Bigg[1+\kappa\frac{\sigma(x)}{M}\Bigg]\Bigg> - \frac{g^2}{2
\pi^2}\int^1_0d\eta\,\eta(1-\eta){\ell n}\Bigg[1 +
\frac{\Box}{M^2}\,\eta(1-\eta)\Bigg] \nonumber\\
&&= Z^{\sigma}_A - \frac{g^2}{2 \pi^2}\int^1_0
d\eta\,\eta(1-\eta){\ell n}\Bigg[1 + \frac{\Box}{M^2}\,\eta(1-\eta)\Bigg],
\end{eqnarray}
and
\begin{eqnarray}\label{label3.16}
&&<M^2_A (\sigma)>= \int {\cal D}\sigma \,M^2_A (\sigma)\,\exp{i\int d^4z\,
{\cal L}_{\rm eff}[\sigma(z)]}=\nonumber\\
&&= \frac{g^2}{2 G_1}-\frac{g^2}{8\pi^2}[J_1(M)+M^2J_2(M)] +
\Bigg<\frac{\kappa g^2}{4 \pi^2}M \sigma(x)
\Bigg[1+\kappa\frac{\sigma(x)}{2M}\Bigg]\Bigg>=\\
&&= \frac{g^2}{2 G_1}-\frac{g^2}{8\pi^2}[J_1(M)+M^2J_2(M)] +
 M^2 Z^{\sigma}_M,\nonumber
\end{eqnarray}
where we have denoted
\begin{eqnarray}\label{label3.17}
Z^{\sigma}_A &=& 1 - \Bigg<\frac{g^2}{6\pi^2}{\ell
n}\Bigg[1+\kappa\frac{\sigma(x)}{M}\Bigg]\Bigg> =\nonumber\\
&=&1 - \int {\cal D}\sigma\,\frac{g^2}{6\pi^2}{\ell
n}\Bigg[1+\kappa\frac{\sigma(x)}{M}\Bigg]\,\exp{i\int d^4z\,
{\cal L}_{\rm eff}[\sigma(z)]},
\end{eqnarray}
and
\begin{eqnarray}\label{label3.18}
Z^{\sigma}_M &=& \Bigg<\frac{\kappa g^2}{2 \pi^2}\,\frac{\sigma(x)}{2 M}
\Bigg[1+\kappa\frac{\sigma(x)}{2M}\Bigg]\Bigg> =\nonumber\\
&=&\int {\cal D}\sigma\,\frac{\kappa g^2}{2 \pi^2}\,\frac{\sigma(x)}{2 M}
\Bigg[1+\kappa\frac{\sigma(x)}{2M}\Bigg]\,\exp{i\int d^4z\,
{\cal L}_{\rm eff}[\sigma(z)]}.
\end{eqnarray}
It is seen that the quantities $Z^{\sigma}_A$ and $Z^{\sigma}_M$ are just
constants, and $Z^{\sigma}_A$ has the meaning of the renormalization
constant
of the wave--function of the $A_{\mu}$--field caused by vacuum fluctuations
of the $\sigma$--field. The vacuum expectation values entering the
constants $Z^{\sigma}_A$ and $Z^{\sigma}_M$ can be represented in the form
of time--ordered products [15]
\begin{eqnarray}\label{label3.19}
Z^{\sigma}_A &=& 1 - \Bigg<\frac{g^2}{6\pi^2}{\ell
n}\Bigg[1+\kappa\frac{\sigma(x)}{M}\Bigg]\Bigg> =\nonumber\\
&=&1 - \int {\cal D}\sigma\,\frac{g^2}{6\pi^2}{\ell
n}\Bigg[1+\kappa\frac{\sigma(x)}{M}\Bigg]\,\exp{i\int d^4z\,
{\cal L}_{\rm eff}[\sigma(z)]}=\nonumber\\
&=&1- \Bigg<0\Bigg|{\rm T}\Bigg(\frac{g^2}{6\pi^2}{\ell
n}\Bigg[1+\kappa\frac{\sigma(x)}{M}\Bigg]\,\exp{i\int d^4z\,
{\cal L}_{\rm int}[\sigma(z)]}\Bigg)\Bigg|0\Bigg>
\end{eqnarray}
and
\begin{eqnarray}\label{label3.20}
Z^{\sigma}_M &=& \Bigg<\frac{\kappa g^2}{2 \pi^2}\,\frac{\sigma(x)}{2 M}
\Bigg[1+\kappa\frac{\sigma(x)}{2M}\Bigg]\Bigg> =\nonumber\\
&=&\int {\cal D}\sigma\,\frac{\kappa g^2}{2 \pi^2}\,\frac{\sigma(x)}{2 M}
\Bigg[1+\kappa\frac{\sigma(x)}{2M}\Bigg]\,\exp{i\int d^4z\,
{\cal L}_{\rm eff}[\sigma(z)]}=\nonumber\\
&=&\Bigg<0\Bigg|{\rm T}\Bigg(\frac{\kappa g^2}{2 \pi^2}\,
\frac{\sigma(x)}{2 M} \Bigg[1+\kappa\frac{\sigma(x)}{2M}\Bigg]\,
\exp{i\int d^4z\,{\cal L}_{\rm int}[\sigma(z)]}\Bigg)\Bigg|0\Bigg>,
\end{eqnarray}
where ${\cal L}_{\rm int}[\sigma(z)]$ reads
\begin{eqnarray}\label{label3.21}
&&{\cal L}_{\rm int}[\sigma(x)] = \frac{1}{2} \partial_{\mu} \sigma(x)
\partial^{\mu} \sigma(x) - \frac{1}{2} M^2_{\sigma} \sigma^2(x)
\Bigg[1 + \kappa \frac{\sigma(x)}{M_{\sigma}}\Bigg]^2 -
V_{\rm eff}[\sigma(x)] -\nonumber\\
&&-\frac{1}{2} \partial_{\mu} \sigma(x) \partial^{\mu} \sigma(x) +
\frac{1}{2} M^2_{\sigma} \sigma^2(x) = - \kappa\,M_{\sigma}\,
\sigma^3(x) - \frac{1}{2}\,\sigma^4(x) - V_{\rm eff}[\sigma(x)].
\end{eqnarray}
In order to understand the structure of the constants $Z^{\sigma}_A$ and
$Z^{\sigma}_M$ we suggest to calculate them in the one--loop approximation
of the $\sigma$--field exchange. One can expect that the  computation of
the constants $Z_A$ and $Z_M$ accounting for multi--loop contributions
should not change the one--loop result substantially but only provide a
redefinition of parameters.

In the one--loop approximation $Z_A$ and $Z_M$ are given by
\begin{eqnarray}\label{label3.22}
Z^{\sigma}_A &=&1 +
\frac{g^2\kappa^2}{3\pi^2}\,\frac{1}{M^2_{\sigma}}\,<0|\sigma^2(x)|0> = 1 +
\frac{\kappa^2 g^2}{48 \pi^4}
\,\frac{\Delta_1(M_{\sigma})}{M^2_{\sigma}},\nonumber\\
Z^{\sigma}_M &=&
\frac{g^2\kappa^2}{2\pi^2}\,\frac{1}{M^2_{\sigma}}\,<0|\sigma^2(x)|0> =
\frac{\kappa^2 g^2}{32\pi^4} \,\frac{\Delta_1(M_{\sigma})}{M^2_{\sigma}},
\end{eqnarray}
where $\Delta_1(M_{\sigma})$ is a quadratically divergent momentum integral
\begin{eqnarray}\label{label3.23}
\Delta_1(M_{\sigma}) =\int\frac{d^4k}{\pi^2 i}\frac{1}{M^2_{\sigma} - k^2}\,.
\end{eqnarray}
The magnitude of $\Delta_1(M_{\sigma})$ depends on the regularization
procedure. For example, within dimensional regularization
$\Delta_1(M_{\sigma})$ is negative. The computation of
$\Delta_1(M_{\sigma})$ by means of a cut--off regularization is not
unambiguous. The result of the computation depends on a shift of the
virtual $\sigma$--field momentum [16]. Indeed, one can define
$\Delta_1(M_{\sigma};Q)$ instead of $\Delta_1(M_{\sigma})$ [16]
\begin{eqnarray}\label{label3.24}
\Delta_1(M_{\sigma};Q) =\int\frac{d^4k}{\pi^2 i}
\frac{1}{M^2_{\sigma} - (k + Q)^2}=\Delta_1(M_{\sigma}) +
\frac{1}{4}\,Q^2 ,
\end{eqnarray}
where $Q$ is an arbitrary 4-vector that can be a space--like one, i.e.
$Q^2 < 0$. Thus, $\Delta_1(M_{\sigma})$ is an arbitrary quantity on both
magnitude and sign. Below we fix $\Delta_1(M_{\sigma})$ using the
compositeness condition for the $A_{\mu}$--field.

The momentum integral $\Delta_1(M_{\sigma})$ resembles the quadratically
divergent integral $J_1(M)$ which we encounter for the computation of the
one--electroquark loop diagram defining the vacuum expectation value
$<\bar{\chi}(0)\,\chi(0)>$. However, as $\Delta_1(M_{\sigma})$ is related
to the contribution of one--loop $\sigma$--field exchange diagrams, the
cut--off parameters applied to the regularization of $\Delta_1(M_{\sigma})$
and $J_1(M)$ can arbitrarilly differ on magnitude. This makes
$\Delta_1(M_{\sigma})$ and $J_1(M)$ independent each other. The integral
$J_1(M)$ suffers the same problem as  $\Delta_1(M_{\sigma})$ and  can be
also considered as an arbitrary parameter of the approach fixed by the
gap--equation Eq.(\ref{label1.3}), i.e. $J_1(M) = 2\pi^2 M/G$.

The Lagrangian Eq.(\ref{label3.14}) can be rewritten as follows
\begin{eqnarray}\label{label3.25}
{\cal L}_{\rm eff}[A(x)]&=&- \frac{1}{4} F_{\mu\nu}(x)\,[Z^{\sigma}_A +
\varepsilon_{\rm eff}(\Box)]\, F^{\mu\nu}(x)
\nonumber\\
&&+\frac{1}{2} M^2_{\rm eff} A_{\mu}(x) A^{\mu}(x)
- j^{\mu}(x) A_{\mu}(x),
\end{eqnarray}
where
\begin{eqnarray}\label{label3.26}
\varepsilon_{\rm eff}(\Box) = - \frac{g^2}{2 \pi^2}\int^1_0
d\eta\,\eta(1-\eta){\ell n}\Bigg[1 + \frac{\Box}{M^2}\,\eta(1-\eta)\Bigg]
\end{eqnarray}
and $M^2_{\rm eff} = <M^2_A(\sigma)>$. We should notice that the
renormalization constant $Z^{\sigma}_A$ does not appear in $M^2_{\rm eff}$.
The appearance of $Z^{\sigma}_A$ in the mass term of the $A_{\mu}$--field
can occur only after the scale transformation of the $A_{\mu}$--field,
${\displaystyle \sqrt{Z^{\sigma}_A}}A_{\mu} \to A_{\mu}$, removing
$Z^{\sigma}_A$ from the kinetic term. Since we deal with the effective
theory and do not perform such a scale transformation of the
$A_{\mu}$--field, the constant $Z^{\sigma}_A$
cannot appear in the effective mass $M^2_{\rm eff}$.

Now let us focus on the behaviour of the dielectric constant
$<\varepsilon(\Box,\sigma)> = Z^{\sigma}_A + \varepsilon_{\rm eff}(k^2)$ as
 a function of $Z^{\sigma}_A$. One can see that the confining behaviour of
the dielectric constant, i.e. $<\varepsilon(k^2, \sigma)>\sim k^2$ at
$k^2\to 0$, can be realized only at $Z^{\sigma}_A=0$. The constraint
$Z^{\sigma}_A=0$
is nothing more than the compositeness condition for the $A_{\mu}$--field
[14]. This should imply that in our approach the confining medium can be
realized
only if the vector field $A_{\mu}$ is a composite field [14] with the
structure more complicated than $\bar{\chi}\gamma_{\mu}\chi$ due to the
contribution of the $\sigma$--field fluctuations. Conventionally, the
structure of the
composite $A_{\mu}$--field might be represented like
$\bar{\chi}\gamma_{\mu}\sigma\chi$.

Assuming that the $A_{\mu}$--field is a composite field, we can impose the
compositeness condition $Z^{\sigma}_A=0$:
\begin{eqnarray}\label{label3.27}
Z^{\sigma}_A = 1 + \frac{\kappa^2 g^2}{48
\pi^4}\,\frac{\Delta_1(M_{\sigma})}{M^2_{\sigma}} = 0\,.
\end{eqnarray}
This fixes $\Delta_1(M_{\sigma})$ in terms of the electroquark mass and the
coupling constants determined by one--electroquark loop diagrams:
$\Delta_1(M_{\sigma}) = -192\pi^4M^2/\kappa^2 g^2 = - 128\pi^4
M^2/\kappa^4$, where we have used the relations $M_{\sigma}=2M$ and
$\kappa^2 = 2\,g^2/3$.

The compositeness condition Eq.(\ref{label3.27}) fixes the value of
$Z^{\sigma}_M$ which yields
\begin{eqnarray}\label{label3.28}
M^2_{\rm eff} = \frac{g^2}{2 G_1}-\frac{g^2}{8\,\pi^2}\,
[J_1(M)\,+\,M^2\,J_2(M)] - \frac{3}{2}\,M^2\,
\end{eqnarray}
at $M_{\sigma}=2M$. Of course, the magnitude of $M^2_{\rm eff}$ is arbitrary
due to the arbitrariness of $G_1$. We can set it equal zero, i.e.
\begin{eqnarray}\label{label3.29}
M^2_{\rm eff} = 0\,.
\end{eqnarray}
Thus, further we deal with a massless vector field $A_{\mu}(x)$ coupled to
the external $\psi$--quark current $j_{\mu}(x) = \bar{\psi}(x) \gamma_{\mu}
\psi(x)$.

For the constraint (\ref{label3.27}) the dielectric constant reads
\begin{eqnarray}\label{label3.30}
&&\varepsilon_{\rm eff}(k^2) =\\
&&= - \frac{g^2}{2 \pi^2}\int^1_0 d\eta\,\eta(1-\eta){\ell n}\Bigg[1 -
\frac{k^2}{M^2}\,\eta(1-\eta)\Bigg]=\frac{g^2}{60 \pi^2}\,
\frac{k^2}{M^2} + O\Bigg(\frac{k^4}{M^4}\Bigg)\,.\nonumber
\end{eqnarray}
In the coordinate space--time this yields
\begin{eqnarray}\label{label3.31}
&&\varepsilon_{\rm eff}(\Box) =\\
&&= - \frac{g^2}{2 \pi^2}\int^1_0 d\eta\,\eta(1-\eta){\ell n}\Bigg[1 +
\frac{\Box}{M^2}\,\eta(1-\eta)\Bigg]= - \frac{g^2}{60 \pi^2}\,
\frac{\Box}{M^2} + O\Bigg(\frac{\Box^2}{M^4}\Bigg)\,.\nonumber
\end{eqnarray}
This behaviour of the dielectric constant produces the confining medium
which provides a linearly rising interquark potential at large relative
distances without inclusion of dual Dirac strings.

\section{Linearly rising interquark potential}
\setcounter{equation}{0}

The effective Lagrangian of the $A_{\mu}$--field coupled to the external
$\psi$--quark current $j_{\mu}(x) = \bar{\psi}(x)\gamma_{\mu}\psi(x)$ in a
medium with the dielectric constant $<\varepsilon(\Box,\sigma)>$ defined by
Eq.(\ref{label3.31}) reads
\begin{eqnarray}\label{label4.1}
{\cal L}_{\rm eff}[A_{\mu}(x)] = -\frac{1}{4}\,F_{\mu\nu}(x)\,
\varepsilon_{\rm eff}(\Box)\, F^{\mu\nu}(x) \,-\,j^{\mu}(x)A_{\mu}(x)\,.
\end{eqnarray}
We have dropped the terms of order $O(A^3)$ and higher which do not
contribute to the kinetic term of the $A_{\mu}$--field in the tree
$A_{\mu}$--field exchange approximation which we are keeping to here.

Varing the Lagrangian (\ref{label4.1}) with respect to $A_{\mu}(x)$ we
derive the equation of motion
\begin{eqnarray}\label{label4.2}
\Box\,\varepsilon_{\rm eff}(\Box)A_{\mu}(x)  = j_{\mu}(x) \,.
\end{eqnarray}
The solution of this equation of motion can be represented as follows
\begin{eqnarray}\label{label4.3}
A_{\mu}(x)  = -\,\int\,d^4x^{\prime}\,{\rm G}^{\rm
NT}(x-x^{\prime}\,)\,j_{\mu}(x^{\prime}\,) \,,
\end{eqnarray}
where ${\rm G}^{\rm NT}(x)$ is the Green function of the NT model given by
the momentum representation
\begin{eqnarray}\label{label4.4}
{\rm G}^{\rm NT}(x)\,=\,\int\frac{d^4k}{(2\pi)^4}\,
\frac{1}{ \varepsilon_{\rm eff}(k^2)}\,\frac{e^{-ik\cdot
x}}{k^2}\,=\,\int\frac{d^4k}{(2\pi)^4}\,\frac{\mu^2}{k^4}\,e^{-ik\cdot x}\,,
\end{eqnarray}
where we have denoted $\mu^2 = 60\pi^2M^2/g^2$. For example, the retarded
Green function reads [17]
\begin{eqnarray}\label{label4.5}
{\rm G}^{\rm NT}_{\rm ret}(x)\,=\,\int\frac{d^4k}{(2\pi)^4}\,
\frac{\mu^2}{[(k^0 + i0)^2 - \vec{k}^{\,2}]^2}\,e^{-ik\cdot
x}\,=\,\frac{\mu^2}{8\pi}\theta(t)\theta(x^2)\,,
\end{eqnarray}
where $x^2=t^2-\vec{x}^{\,2}$. The complete set of the Green functions for
the model [1] has been computed in [18].

The effective interquark potential, defined in terms of the Green function
${\rm G}^{\rm NT}(x)$, reads [1]
\begin{eqnarray}\label{label4.6}
V(\vec{r}\,) = - \int^{\infty}_{-\infty}dt{\rm G}^{\rm NT}(t,\vec{r}\,) =
\sigma_{\rm string} r + {\it an~infrared~divergent~constant},
\end{eqnarray}
where $\sigma_{\rm string} = \mu^2/8\pi$ can be identified with the string
tension [3,19]. The string tension, given by the expression
 $\sigma_{\rm string} = \mu^2/8\pi$, has been computed in Ref.[19], where
$1/\mu$ has been identified with the penetration depth of the dual electric
field in a dual superconductor.

\section{Conclusion}

We have shown that in the Abelian NJL model, analoguous the monopole NJL
model [10] and the Abelian version of the technicolour extension of the
standard electroweak model [12], one can realize a medium caused by quantum
fluctuations of the electorquark fields $\chi$, an Abelian analog of
technifermions, and the scalar field $\sigma$, collective $\bar{\chi}\chi$
excitation. The dielectric constant of this medium $\varepsilon(k^2)$
vanishes at large distances, i.e. at $k^2 \to 0$, like $\varepsilon(k^2) =
k^2/8\pi\sigma_{\rm string}$, where $\sigma_{\rm string}$ can be identified
with a string tension. This medium
leads to confinement of the external quark fields $\psi(x)$ if they couple
to the medium and each other via the exchange of the massless composite
vector fields $A_{\mu}(x)$ which are the collective excitations with a
conventional structure $\bar{\chi} \gamma^{\mu} \sigma \chi$. The confining
medium induces a linearly rising interquark potential
$V_{\bar{\psi}\psi}(\vec{r}\,) = \sigma_{\rm string}\,r + C$ without the
inclusion of dual Dirac strings, where the string tension $\sigma_{\rm
string}$ is expressed in terms of the electroquark mass
$M$ and the cut--off $\Lambda$, i.e. $\sigma_{\rm string} =
(5M^2/8\pi)\,J_2(M)$.

The discussions with Prof. H. Narnhofer, Prof. W. Thirring and Prof. V.
Gogohia are appreciated.


\newpage

\begin{thebibliography}{9}
\bibitem{[1]}
H. Narnhofer and W. Thirring,
in Rigorous Methods in Particle Physics, eds. S. Ciulli, F. Scheck and W.
Thirring, Springer Tracts in Modern Physics Vol. 119 (1990) 1.
\bibitem{[2]}
M. Baker, J. Ball and F. Zachariasen,
Phys. Rev. D 41 (1990) 2612; Phys. Rep. 209 (1991) 73.
\bibitem{[3]}
A. N. Ivanov, N. I. Troitskaya, M. Faber, M. Schaler and M. Nagy, Phys.
Lett. B336 (1994) 555; Nuovo Cim. A 107 (1994) 1667;
A. N. Ivanov, N. I. Troitskaya and M. Faber,
Nuovo. Cim. A 108 (1995) 613.
\bibitem{[4]}
Y. Nambu and G. Jona--Lasinio,
Phys. Rev. 122 (1961) 345; ibid. 124 (1961) 246.
\bibitem{[5]}
T. Eguchi,
Phys. Rev. D 14 (1976) 2755;
K. Kikkawa,
Progr. Theor. Phys. 56 (1976) 947;
H. Kleinert,
Proc. of Int. Summer School of Subnuclear
Physics, Erice 1976, Ed. A.Zichichi, p.289.
\bibitem{[6]}
T. Hatsuda and T. Kumihiro,
Proc. Theor. Phys. 74 (1985) 765; Phys. Lett. B 98 (1987) 126;
T. Kumihiro and T. Hatsuda,
Phys. Lett. B 206 (1988) 385.
\bibitem{[7]}
S. Klint, M. Lutz, V. Vogl and W. Weise,
Nucl. Phys. A 516 (1990) 429; 469 and references therein.
\bibitem{[8]}
A. N. Ivanov, M. Nagy and N. I. Troitskaya,
Int. J. Mod. Phys. A 7 (1992) 7305;
A. N. Ivanov, Int. J. Mod. Phys. A 8 (1993) 853;
A. N. Ivanov, N. I. Troitskaya and M. Nagy,
Int. J. Mod. Phys. A 8 (1993) 2027; ibid. A 8 (1993) 3425.
\bibitem{[9]}
A. N. Ivanov, N. I. Troitskaya and M. Nagy,
Phys. Lett. B308 (1993) 111; ibid. B 295 (1992) 308.
A. N. Ivanov and N. I. Troitskaya,
Nuovo Cim. A 108 (1995) 555.
\bibitem{[10]}
M. Faber, A. N. Ivanov , W. Kainz and N. I. Troitskaya,
Z. Phys. C74 (1997) 721; Phys. Lett. B 386 (1996) 198.
\bibitem{[11]}
J. Bardeen, L. N. Cooper and J. R. Schrieffer,
Phys. Rev. 106 (1957) 162; ibid. 108 (1957) 1175.
\bibitem{[12]}
S. Weinberg,
Phys. Rev. D19 (1979) 1277;
L. Susskind,
Phys. Rev. D20 (1979) 2619.
\bibitem{[13]}
C. Itzykson and J.--B. Zuber, in {\it Quantum Field  Theory}
(McDraw--Hill), 1980.
\bibitem{[14]}
K. Hayashi, M. Hirayama, T. Muta, N. Seto and T. Shirafuji,
Fortschritte der Physik, 15 (1967) 625;
G. V. Efimov and M. A. Ivanov, in {\it Quark confinement model of hadrons},
Institute of Physics Publishing Bristol and Philadelphia, London 1993, p.21;
G. V. Efimov, {\it On bound states in quantum field theory}, Bogoliubov
Laboratory of Theoretical Physics, JINR, 141980 Dubna, Russia,
hep--ph/9607425 25 July 1996, 33p.
\bibitem{[15]}
N. N. Bogoliubov and D. V. Shirkov,
in {\it Introduction to the theory of quantized fields}, Interscience
Publishers, Inc., New York Interscience Publishers Ltd., London, Chapter
VII, 1959.
\bibitem{[16]}
I. S. Gertsein and R. Jackiw,
Phys. Rev. 181 (1969) 1955.
\bibitem{[17]}
H. Narnhofer and W.~Thirring,
Phys. Lett. B 76 (1978) 428.
\bibitem{[18]}
M. Faber, A. N. Ivanov and N. I. Troitskaya,
{\it Green functions in the Narnhofer--Thirring model},
Preprint of Institut f\"ur Kernphysik Technische Universit\"at Wien,
February, 1996.
\bibitem{[19]}
M. Faber, W. Kainz, A. N. Ivanov and N. I. Troitskaya,
Phys. Lett. B 344  (1995) 143.
\end{thebibliography}
\end{document}